\documentstyle[prl,aps,multicol]{revtex}

\begin{document}
\draft

\title{ Quantum disorder versus order-out-of-disorder in the Kugel-Khomskii
        model }

\author { Louis Felix Feiner \cite{LFF} }
\address{ Philips Research Laboratories, Prof. Holstlaan 4,
          NL-5656 AA Eindhoven, The Netherlands, and \\
          Institute for Theoretical Physics, Utrecht University,
          Princetonplein 5, NL-3584 CC Utrecht, The Netherlands }
\author { Andrzej M. Ole\'{s} }
\address{ Max-Planck-Institut f\"ur Festk\"orperforschung,
          Heisenbergstrasse 1, D-70569 Stuttgart,
          Federal Republic of Germany, and 
          Institute of Physics, Jagellonian University, Reymonta 4,
          PL-30059 Krak\'ow, Poland \cite{AMO} }
\author { Jan Zaanen }
\address{ Lorentz Institute for Theoretical Physics, Leiden University,
          P.O.B. 9506, NL-2300 RA Leiden, The Netherlands }
\date{May 15, 1998}
\maketitle

\begin{abstract}
The Kugel-Khomskii model, the simplest model for orbital degenerate 
magnetic insulators, exhibits a zero temperature degeneracy in the 
classical limit which could cause genuine quantum disorder.
Khaliullin and Oudovenko [Phys. Rev. B {\bf 56}, R14\ 243 (1997)]
suggested recently that instead a particular classical state could 
be stabilized by quantum fluctuations. Here we compare their approach 
with standard random phase approximation and show that it strongly 
underestimates the strength of the quantum fluctuations,
shedding doubts on the survival of any classical state.
\end{abstract}


\begin{multicols}{2} 

Motivated by the developments in the manganites, the interest in the role
of orbital degeneracy in strongly correlated systems revived.
A classic model in this context is the Kugel-Khomskii model \cite{Kug73},
believed to be realistic \cite{Lie95} for KCuF$_3$ and related systems 
(one hole per site, degeneracy of the $e_g$ orbitals). We recently 
discovered that this model poses a rather fundamental problem \cite{Fei97}:
in the classical limit a point exists
in the space of physical parameters where its ground state becomes infinitely
degenerate, due to a novel dynamical frustration mechanism. This classical
degeneracy is lifted on the quantum level, and by analyzing valence-bond
type variational states we arrived at the suggestion that the ground state
for $S=1/2$ might well be an incompressible spin fluid. In a follow up
paper, Khaliullin and Oudovenko \cite{Kha97} suggested that instead
the quantum fluctuations act to single out one particular classical state
(the one with N\'eel order and $d_{3z^2-r^2}$ orbitals occupied by holes)
over all others by an order-out-of disorder mechanism: the classical 
degeneracy is lifted by the differing strength of the fluctuations around 
the various classical states but these fluctuations are not severe enough to 
destroy the classical N\'eel order completely. Their suggestion was based on 
a particular decoupling scheme and in this communication we will demonstrate 
that for rather simple reasons this decoupling scheme implies a serious 
underestimation of the strength of the fluctuations, shedding serious doubts 
on the possibility that classical order survives after all.

The Kugel-Khomskii model describes a three-di\-men\-sional (3D) cubic 
Mott-Hubbard insulator with a single hole/electron in $e_g$ 
orbitals ($x^2-y^2\sim | x \rangle$, $3z^2-1\sim | z \rangle$), possessing, 
in the absence of virtual hoppings, orbital degeneracy in addition to the 
standard spin degeneracy. Its minimal version is given by ($J=t^2/U$ being 
the antiferromagnetic (AF) superexchange, with $t$ the hopping element
between $|z\rangle$ orbitals along the $c$-axis) \cite{Kug73,Fei97},
\begin{eqnarray}
\label{som}
H &=& J \sum_{\langle ij\rangle} \left[4(\vec{S}_i\cdot\vec{S}_j )
  (\tau^{\alpha}_i-\case{1}{2}) (\tau^{\alpha}_j-\case{1}{2})     \right.
                                   \nonumber \\
& & \hskip .8cm + \left.
  (\tau^{\alpha}_i+\case{1}{2}) (\tau^{\alpha}_j+\case{1}{2}) - 1 \right]
  - E_z \sum_i \tau^c_i,
\end{eqnarray}
where $E_z$ is the energy splitting between the $e_g$ orbitals, acting as
a "magnetic field" for the orbital pseudo-spins. It is used to investigate 
the system when it approaches the degeneracy point $E_z=0$.
The spin operators $\vec{S}_i$ are $S=1/2$ spins, while the orbital degrees
of freedom are described by $(2\times 2)$ matrices in the pseudospin space,
\begin{equation}
\label{orbop}
\tau^{a(b)}_i = \case{1}{4}( -\sigma^z_i \pm\sqrt{3}\sigma^x_i ), \hskip .7cm
\tau^c_i      = \case{1}{2}   \sigma^z_i,
\end{equation}
and $\alpha$ selects the cubic axis ($a,b$, or $c$) that corresponds to
the orientation of the bond $\langle ij\rangle$.
The $\sigma$'s are Pauli matrices acting on the orbital pseudo-spins
$|x\rangle ={\scriptsize\left( \begin{array}{c} 1\\ 0\end{array}\right)},\;
 |z\rangle ={\scriptsize\left( \begin{array}{c} 0\\ 1\end{array}\right)}$.
Hence, a Heisenberg model for the spins is coupled into an orbital problem.
Here we ignore the (physically important) multiplet splittings due to a 
finite value of the atomic Hund's rule coupling ($J_H$), and focus on
the special point $E_z, J_H \rightarrow 0$,
contained in model Eq. (\ref{som}): it is easy to see\cite{Fei97} that
in the {\em classical} limit the system is dynamically frustrated and an
infinity of classical phases become degenerate at zero temperature. This
degeneracy is lifted on the quantum level and one expects quantum effects
to take over at this point itself, as well as in its direct vicinity 
\cite{Fei97}, in analogy to what seems established in geometrically
frustrated spin models \cite{Pre88}. If a disordered state would be 
stabilized by quantum effects, orbital degeneracy could be added to the 
list of mechanisms leading to a spin-liquid, such as the frustrated 
$J_1-J_2$ Heisenberg antiferromagnet (HAF) \cite{Pre88}, the bilayer HAF
\cite{Mil94}, and two-dimensional (2D) lattices with
a reduced number of magnetic bonds, as realized in CaV$_4$O$_9$ \cite{Ued96}.

Quite generally, the transverse modes \cite{Fei97,Kha97,Ish96} may be
calculated starting from the equations of motion \cite{Hal72},
\begin{eqnarray}
\label{gfs}
E\langle\langle {\cal S}_{i}^+|...\rangle\rangle &=&
{1\over 2\pi}\langle [{\cal S}_{i}^+,...]\rangle +
 \langle\langle [{\cal S}_{i}^+,H]|...\rangle\rangle,     \\
\label{gfo}
E\langle\langle {\cal K}_{i}^+|...\rangle\rangle &=&
{1\over 2\pi}\langle [{\cal K}_{i}^+,...]\rangle +
 \langle\langle [{\cal K}_{i}^+,H]|...\rangle\rangle,      
\end{eqnarray}
and using a generalization of the LSW theory.
Here ${\cal S}_i^{+}$ is either
$S_i^+$ or $\tilde{S}_i^+ \equiv S_i^+ \sigma_i^z$,
while ${\cal K}_i^{+}$ is either
$K_i^{++} \equiv S_i^+ \sigma_i^+$ or
$K_i^{+-} \equiv S_i^+ \sigma_i^-$. The first pair of
Green functions stands for {\em spin-wave\/} (SW) excitations, while the
second pair describes mixed {\em spin-and-orbital-wave\/} (SOW) excitations.
Similarly a longitudinal mode is given by
\begin{equation}
E\langle\langle {\sigma}_{i}^+|...\rangle\rangle =
{1\over 2\pi}\langle [{\sigma}_{i}^+,...]\rangle +
 \langle\langle [{\sigma}_{i}^+,H]|...\rangle\rangle,
\label{gfl}
\end{equation}
where the Green function describes a {\em purely orbital\/} excitation.
At each site the full set of local operators describing these excitations
constitutes a $so(4)$ Lie algebra. Even though the spin-wave operators form 
a subalgebra, as seen from the familiar $su(2)$ commutators together with
the additional commutators
\begin{equation}
{[S_i^+,S_i^z \sigma_i^z]}=- \tilde{S}_i^+,   \hskip .8cm
{[\tilde{S}_i^+,S_i^z \sigma_i^z]}=- S_i^+ ,
\label{com1}
\end{equation}
while the same holds for the spin-and-orbital operators,
\begin{equation}
{[K_i^{+\pm},S_i^z]}=- K_i^{+\pm},   \hskip .4cm
{[K_i^{+\pm},K_i^{-\mp}]}= 4 S_i^z \pm 2 \sigma_i^z,
\label{com2}
\end{equation}
one needs with Hamiltonian (\ref{som}) for the calculation of
the SW and SOW excitations also commutators like
\begin{equation}
\label{com3}
{[S_i^+,S_i^z \sigma_i^{\pm}]}=- K_i^{+\pm},   \hskip .5cm
{[K_i^{+\pm},S_i^z \sigma_i^{\mp}]} = -2 S_i^+ .
\end{equation}
Clearly, {\it the SOWs cannot be separated from the SWs\/},
and one has to solve simultaneously Eqs. (\ref{gfs}) and (\ref{gfo}).

The random-phase approximation (RPA) for spinlike operators linearizes 
the equations of motion by the familiar decoupling procedure \cite{Hal72},
\begin{equation}
\langle\langle {\cal A}_i {\cal B}_j|...\rangle\rangle \simeq
\langle {\cal A}_i\rangle \langle\langle {\cal B}_j|...\rangle\rangle +
\langle {\cal B}_j\rangle \langle\langle {\cal A}_i|...\rangle\rangle .
\label{deco}
\end{equation}
It is crucial that the decoupled operators ${\cal A}_i$ and ${\cal B}_j$
have {\em different\/} site indices, so that this procedure does not violate 
the local Lie-algebraic structure of the commutation rules 
(\ref{com1}-\ref{com3}). In the N\'eel-type AF phases with either $|x\rangle$ 
(AFxx) or $|z\rangle$ (AFzz) orbitals occupied, one now finds after Fourier 
transformation and using the nonzero expectation values of 
$S_j^z$, $\sigma_j^z$, and $S_j^z\sigma_j^z$ operators, two excitations 
($\alpha={\rm x,z}$ for AFxx and AFzz, respectively),
\begin{eqnarray}
\label{omega}
[\omega_{\vec k}^{(n)}]^2&=&
      \case{1}{2}J^2\left(\lambda_{\alpha}^2+\tau_{\alpha}^2
      -Q_{\alpha\vec k}^2-R_{\vec k}^2-2P_{\alpha\vec k}^2\right)
                                                           \nonumber \\
&\pm &\case{1}{2}J^2\left[ (\lambda_{\alpha}^2-\tau_{\alpha}^2)^2
      -2(\lambda_{\alpha}^2-\tau_{\alpha}^2)(Q_{\alpha\vec k}^2-R_{\vec k}^2)
      \right.                                              \nonumber \\
&-&\left. 4(\lambda_{\alpha}-\tau_{\alpha})^2P_{\alpha\vec k}^2
      +(Q_{\alpha\vec k}^2+R_{\vec k}^2+2P_{\alpha\vec k}^2)^2 \right.
                                                           \nonumber \\
&-&\left. 4(Q_{\alpha\vec k}R_{\vec k}-P_{\alpha\vec k}^2)^2\right]^{1/2}.
\end{eqnarray}
The orbital dependence enters the ${\vec k}$-independent field,
\begin{equation}
\label{lambda}
\lambda_{\rm x(z)}=\case{9}{2},\hskip 1.0cm
\tau_{\rm x(z)}=\case{3}{2}\pm\varepsilon_z,
\end{equation}
with $\varepsilon_z=E_z/J$, and the dispersion is given by,
\begin{eqnarray}
\label{bothq}
Q_{{\rm x}\vec k}&=&\case{9}{2}\gamma_{+}(\vec k),            \hskip 0.9cm
Q_{{\rm z}\vec k} = \case{1}{2}\gamma_{+}(\vec k)+4\gamma_{z}(\vec k),  \\
\label{bothp}
P_{{\rm x}\vec k}&=&\case{3}{2}\sqrt{3}\gamma_{-}(\vec k),    \hskip 0.5cm
P_{{\rm z}\vec k} = \case{1}{2}\sqrt{3}\gamma_{-}(\vec k),    \\
& & \hskip 1.4cm  R_{\vec k}=\case{3}{2}\gamma_{+}(\vec k),
\label{oner}
\end{eqnarray}
with $\gamma_{\pm}(\vec k)=\frac{1}{2}(\cos k_x\pm\cos k_y)$ and
     $\gamma_{z}(\vec k)  =           \cos k_z$.

The dispersions of SW and SOW are shown in Fig. \ref{realspectra}. It is
straightforward to verify that the SW dispersion is $9J/2$, given for the
AFxx phase by the superexchange of $9J/4$ between $|x\rangle$ orbitals in
the $(a,b)$-planes, and for the AFzz phase by strong interactions of $4J$
along the $c$-axis and weak superexchange of $J/4$ in the $(a,b)$-planes.
In both phases one finds that the coupling between the modes due to the
$P_{\alpha\vec k}\sim\gamma_-(\vec k)$ term is strong, and the excitations 
have pure character only in the planes of $\gamma_{-}(\vec k)=0$, as seen 
along $\Gamma-L(K)$ lines. In particular, this coupling increases along the
$\Gamma-X$ direction, and precisely compensates the dispersion due to the
orbital dynamics $\sim\gamma_+(\vec k)$. This results in a {\em soft mode\/}
$\omega_{\vec k}^{(1)}=0$ along the $\Gamma-X(Y)$ direction in both AF
phases. As we have shown before \cite{Fei97}, finite masses are found in the 
directions perpendicular to the soft mode lines, which gives a logarithmic 
divergence of the quantum correction to the order parameter,
$\langle\delta S^z\rangle\sim\ln\Delta_i$, with $\Delta_i\to 0$ for $E_z\to 0$.

Khaliullin and Oudovenko \cite{Kha97} calculate instead a SW and a {\em 
longitudinal excitation\/} $\langle\langle \sigma_{i}^+|...\rangle\rangle$ 
(\ref{gfl}) first, and then include the effect of orbital fluctuations in 
the transverse channel (our SOW) in a perturbative way. 
Their selfconsistent calculation gives a finite energy and weak dispersion 
of the orbital mode (in the present RPA approach the orbital excitation 
is found at $\omega=0$). This approach violates the commutation relations 
in the Lie algebra (\ref{com1}-\ref{com3}), and only for this reason the SW
and SOW excitations become independent from each other. In the present RPA 
language it implies that composite spin-and-orbital operators,
$S_i^{\alpha} \sigma_i^{\beta}$, are factorized 
into {\em independent products\/} of spin ($S_i^{\alpha}$) and orbital
($\sigma_i^{\beta}$) operators separately, and the commutators given by Eqs.
(\ref{com1}-\ref{com3}) effectively either vanish, {\it e.g.\/} 
$[S_i^+,S_i^z         \sigma_i^{\pm}       ]\mapsto 
 [S_i^+,S_i^z]\langle \sigma_i^{\pm}\rangle = 0$,
or give a different result, {\it e.g.\/} 
$[S_i^+,S_i^z\sigma_i^z]\mapsto [S_i^+,S_i^z]\langle\sigma_i^z\rangle=
 -S_i^+\langle\sigma_i^z\rangle$.
We call this procedure the SW+SOW scheme; it is formally equivalent to 
assuming $P_{\alpha\vec k}=0$ in Eq. (\ref{omega}). The SW modes depend 
now solely on the actual magnetic interactions, while the SOW modes are 
identical in the two phases and the soft mode behaviour is absent 
(Fig. \ref{poorspectra}).
However, the SOW mode is gapless at the $\Gamma$ point as a consequence of 
the rotational invariance of the classical AF order when the occupied 
orbitals are rotated between $|x\rangle$ and $|z\rangle$.

We calculated the order parameter $\langle S^z\rangle$ in both AF phases
including quantum corrections using a generalized RPA approach
which leads to the identity,
\begin{equation}
\label{reno}
\langle S_i^z\rangle_{\rm RPA}= \frac{1}{2}-\langle S_i^-S_i^+\rangle
                    -\frac{1}{2}\langle S_i^-\sigma_i^-S_i^+\sigma_i^+\rangle,
\end{equation}
where $i\in A$, and $A$ is the $\uparrow$-spin sublattice.
The identity (\protect{\ref{reno}}) follows from the expansion of the 
$S_i^z$ operator in the $so(4)$ algebra and replaces the $su(2)$ relation
$\langle S_i^z\rangle=\frac{1}{2}-\langle S_i^-S_i^+\rangle$, familiar from 
the Heisenberg model. It includes the renormalization due to {\em both\/} 
transverse modes in the spin-orbital model (\ref{som}). Similarly the orbital 
occupancy is renormalized by $\langle \sigma^{-}_i \sigma^{+}_i \rangle$ 
fluctuations due to the longitudinal mode. The correlation functions are 
found from the respective Green functions \cite{Hal72},
\begin{equation}
\label{calav}
\langle {\cal A}_i{\cal B}_i\rangle=\int\limits_{-\infty}^0d\omega
\left(\frac{1}{N}\sum_{\vec k}2\Im
\langle\langle {\cal B}_{\vec k}|
               {\cal A}_{\vec k}\rangle\rangle_{\omega-i\epsilon}\right).
\end{equation}
Eq. (\ref{reno}) reproduces the result for the 2D HAF,
$\langle S_i^z\rangle\simeq 0.303$, in the limit of $E_z\to+\infty$, while
$\langle S_i^z\rangle\simeq 0.251$ for the strongly anisotropic 3D HAF at
$E_z\to-\infty$. The values of $\langle S_i^z\rangle$ are, however, strongly
reduced when the degeneracy point ($E_z=0$) is approached (Fig. \ref{sz}),
and the quantum corrections {\em overshoot\/} the mean-field value
$\langle S_i^z\rangle_{\rm MF}$ for $-0.04<E_z/J<0.30$, and diverge at
$E_z=0$. In contrast, these corrections are much reduced within the SW+SOW
scheme, and the divergence at $E_z$ is removed
($\langle S_i^z\rangle\simeq 0.05$ in both phases). This is qualitatively
equivalent to the results of Ref. \onlinecite{Kha97}, where the
renormalization of $\langle S_i^z\rangle$ due to the SOW was included only
perturbatively, and a value $0.191$ was found in the AFzz phase.
This somewhat smaller quantum correction results from the finite
gap in the orbital excitation.

Further evidence that the stability of the LRO phases is overstimated in
Ref. \onlinecite{Kha97} comes from energy calculations. 
For convenience we define the ground state energy per site as a quantum
correction beyond the mean-field value,
\begin{equation}
\label{egs}
E=\frac{1}{N}\langle H\rangle+E_z\langle \tau_{i}^c\rangle+3J.
\end{equation}
A simple estimation at $E_z=0$ using the Bethe ansatz solution for a
disordered one-dimensional (1D) chain along the $c$-axis, and no magnetic
correlations in the $(a,b)$-planes gives $E=-0.648J$ \cite{Fei97}, while a
somewhat better energy of $-0.656J$ was obtained using plaquette valence 
bond (PVB) states either with singlets alternating along the $a$- and 
$b$-axis in the $(a,b)$-planes (PVBA phase), or with single planes of such 
alternating singlets interlayered with two planes of singlets along the 
$c$-axis (PVBI phase), as explained in Ref. \onlinecite{Fei97}.

For the LRO phases, in spite of the divergent correction to the order
parameter, an energy can still be estimated using the RPA corrections for 
the symmetry-broken state. Here these estimates starting from the states 
with LRO give lower energies than the above simple estimates for the VB 
states. This is not surprising, as it is known that improved VB wave
functions that include the resonance between spin singlets lead typically 
to large energy gains, but are difficult to treat already in spin
models \cite{Lia88}. First, within the generalized RPA approach one finds 
the largest quantum corrections in the AFxx phase (Fig. \ref{energy}). This 
shows that the AFxx phase is `more unstable' against disorder, in agreement 
with intuition and with Ref. \onlinecite{Kha97}. We believe that the lowest 
energy $-0.896J$ obtained in the AFxx phase at $E_z=0$ comes close to the 
true ground state. This is consistent with the experience with the 1D HAF, 
where one finds an energy of $-0.429J$ using the LSW theory, which is only 
3.2\% above the exact value $-0.443J$. We note that the energy obtained 
within the simplified SW+SOW approach is much higher, even above that of the 
disordered phases (PVB states). In contrast, the SW+SOW approach gives for 
the AFzz phase an energy somewhat lower than that of the PVB states, and our 
value of $E$ differs only by $0.005J$ from that reported by Khaliullin and 
Oudovenko in their scheme [Table I]. This indicates the qualitative 
similarity of these two approximations in treating the quantum fluctuations 
related to simultaneous spin and orbital flips (SOW excitations) -- in both 
cases the effect of such fluctuations is severely underestimated.

Summarizing, the results presented in Ref. \onlinecite{Kha97} are
inconclusive, as their approximation violates the $so(4)$ dynamical algebra
describing the microscopic excitations. In contrast to the result of the
perturbative treatment of Ref. \onlinecite{Kha97}, the RPA calculation 
yields an unstable AFzz (and AFxx) phase at orbital degeneracy, as also 
found in spin systems \cite{Pre88}. As the LSW theory performs quite well 
in simple spin systems with $S=1/2$ \cite{Cha88}, this strongly suggests 
that the ground state of the Kugel-Khomskii model is a spin-liquid. To our 
knowledge, the present case is unique in the sense that singlets arranged 
in a 3D valence bond solid (PVB states) allow a lower energy than that of 
a classical state.  However, it might well be that the final verdict on 
these matters has to wait for the systematic approach to the quantization 
of classically frustrated problems, which is still to be invented.

We acknowledge the support by the Committee of Scientific
Research (KBN) of Poland, Project No. 2 P03B 175 14 (AMO), and by the
Dutch Academy of Sciences (KNAW) (JZ).


\narrowtext


\begin{table}
\caption{
Ground state energy $E$, in units of $J$, as obtained for the Kugel-Khomskii
model using the full RPA (RPA) and decoupled SW and SOW excitations (SW+SOW),
compared with the energy found in Ref. \protect\onlinecite{Kha97}.  }
\label{table1}
\begin{tabular}{ccccccc}
   method                         &  AFzz phase  &  AFxx phase  \\
\tableline
    RPA                           &    -0.745    &    -0.896    \\
  SW+SOW                          &    -0.685    &    -0.474    \\
Ref. \protect{\onlinecite{Kha97}} &    -0.690    &       -      \\
\end{tabular}
\end{table}


\begin{figure}
\caption
{Transverse excitations $\omega_k/J$ for the Kugel-Khomskii model at orbital
degeneracy ($E_z=0$) within RPA for the AFzz (top) and AFxx (bottom) phases
in the $fcc$ (AFzz) Brillouin zone. Strong coupling between the (SW and SOW)
modes results in a soft mode along the $\Gamma-X(Y)$ direction.}
\label{realspectra}
\end{figure}

\begin{figure}
\caption
{The same as in Fig. \protect{\ref{realspectra}}, but within the simplified
SW+SOW scheme; the SOW dispersion is $1.5J$.}
\label{poorspectra}
\end{figure}

\begin{figure}
\caption
{Order parameters $\langle S^z_i\rangle$ for AFzz (left) and AFxx (right)
phases as functions of $E_z/J$ using: full RPA (full lines), and SW+SWO
scheme (dashed lines). The horizontal lines show the limits found
at $E_z/J\to -\infty$ (dashed line), and at $E_z/J\to \infty$ (2D HAF,
dashed-dotted line).}
\label{sz}
\end{figure}

\begin{figure}
\caption
{Ground state energies $E$ of the AFzz (left) and AFxx (right) phases as
functions of $E_z/J$, obtained using RPA (full lines), and SW+SWO scheme
(dashed lines).}
\label{energy}
\end{figure}

\end{multicols} 

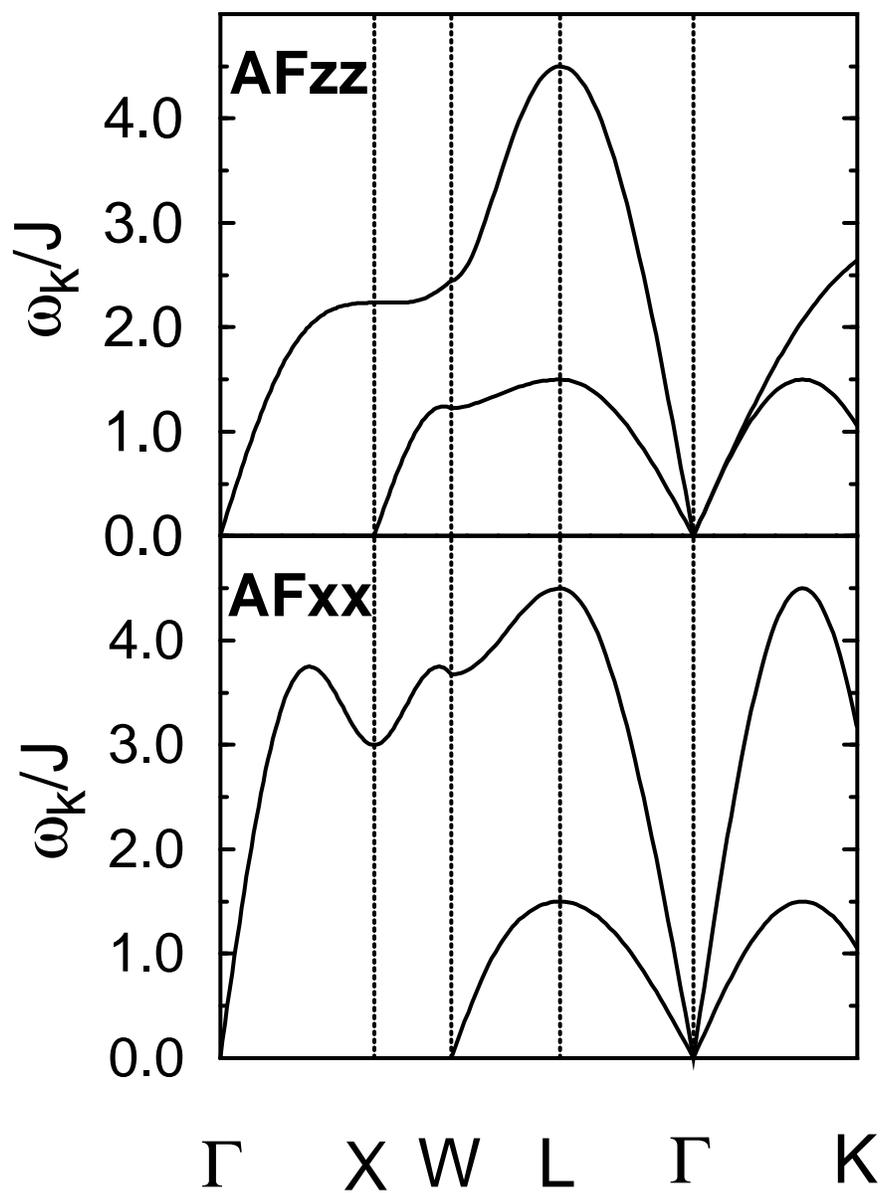
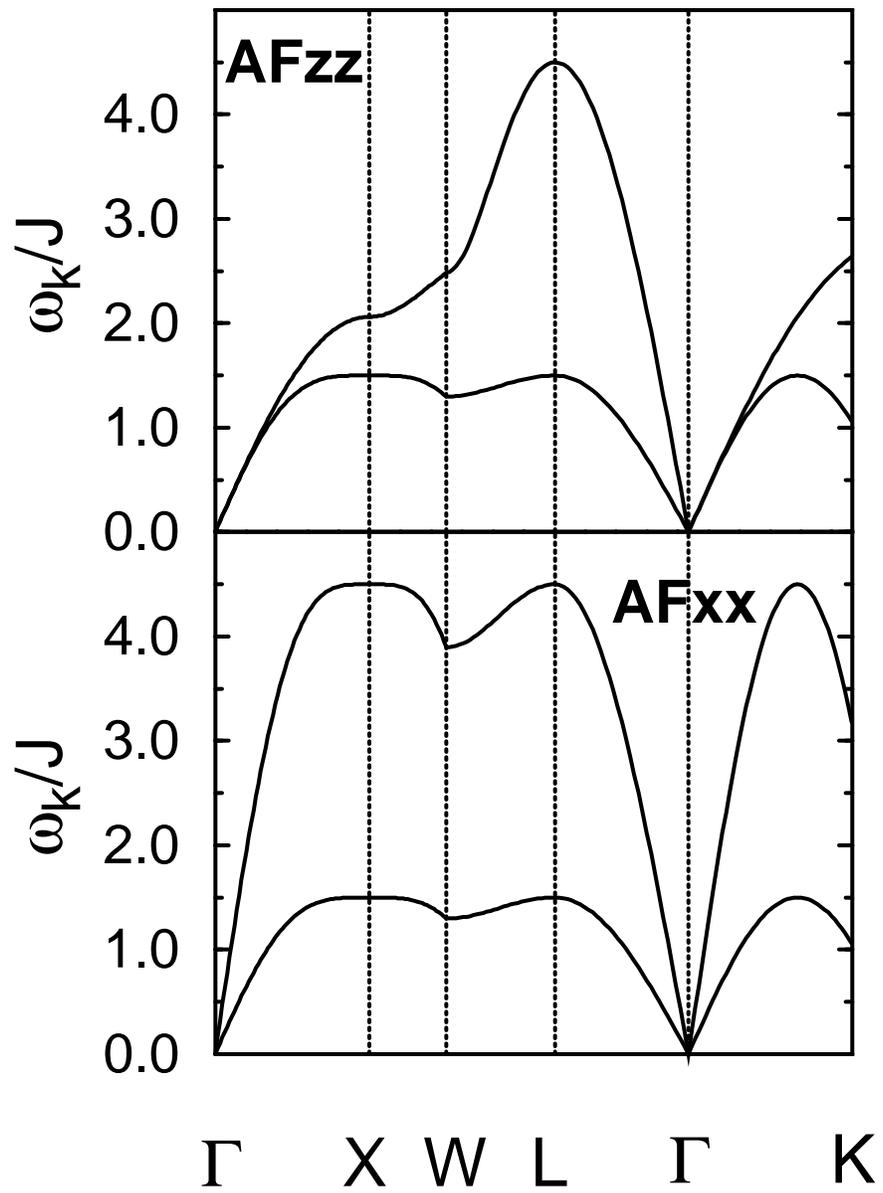
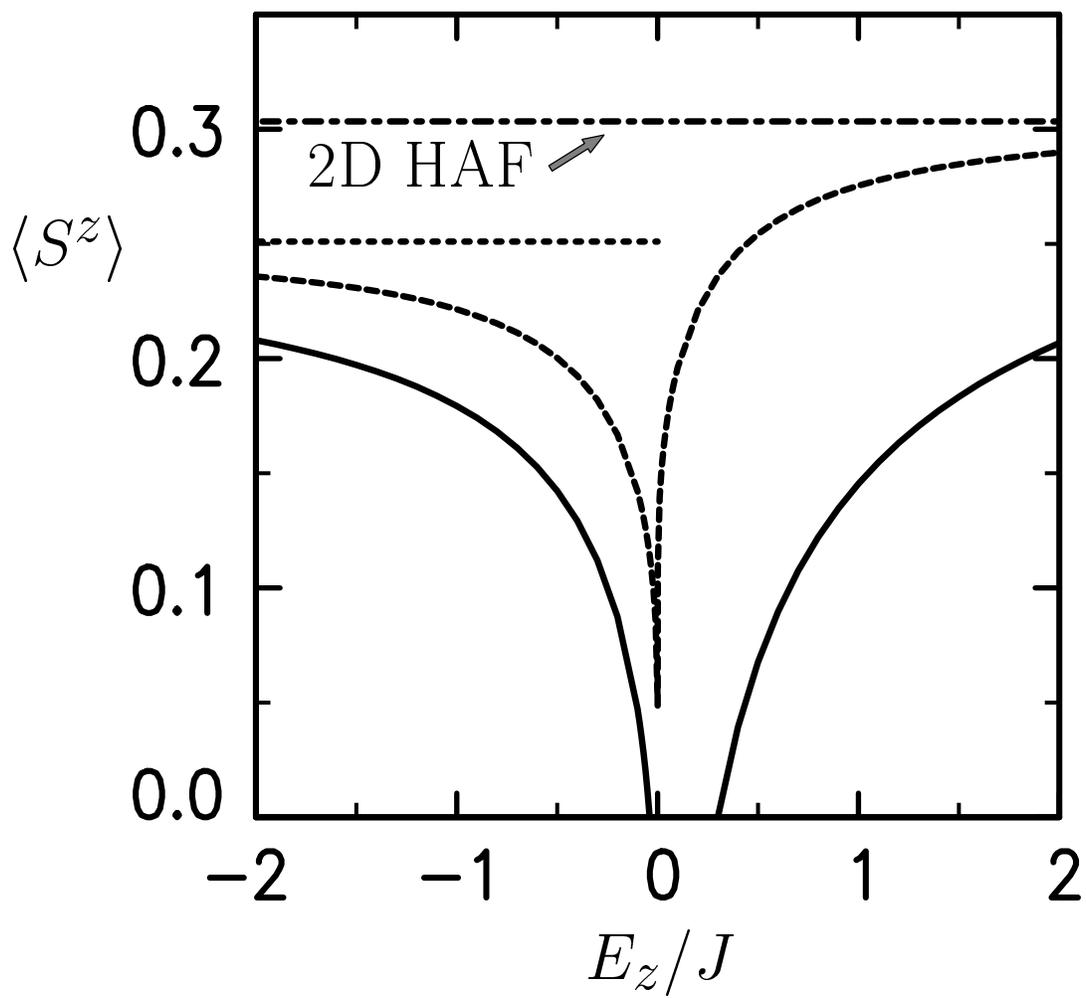
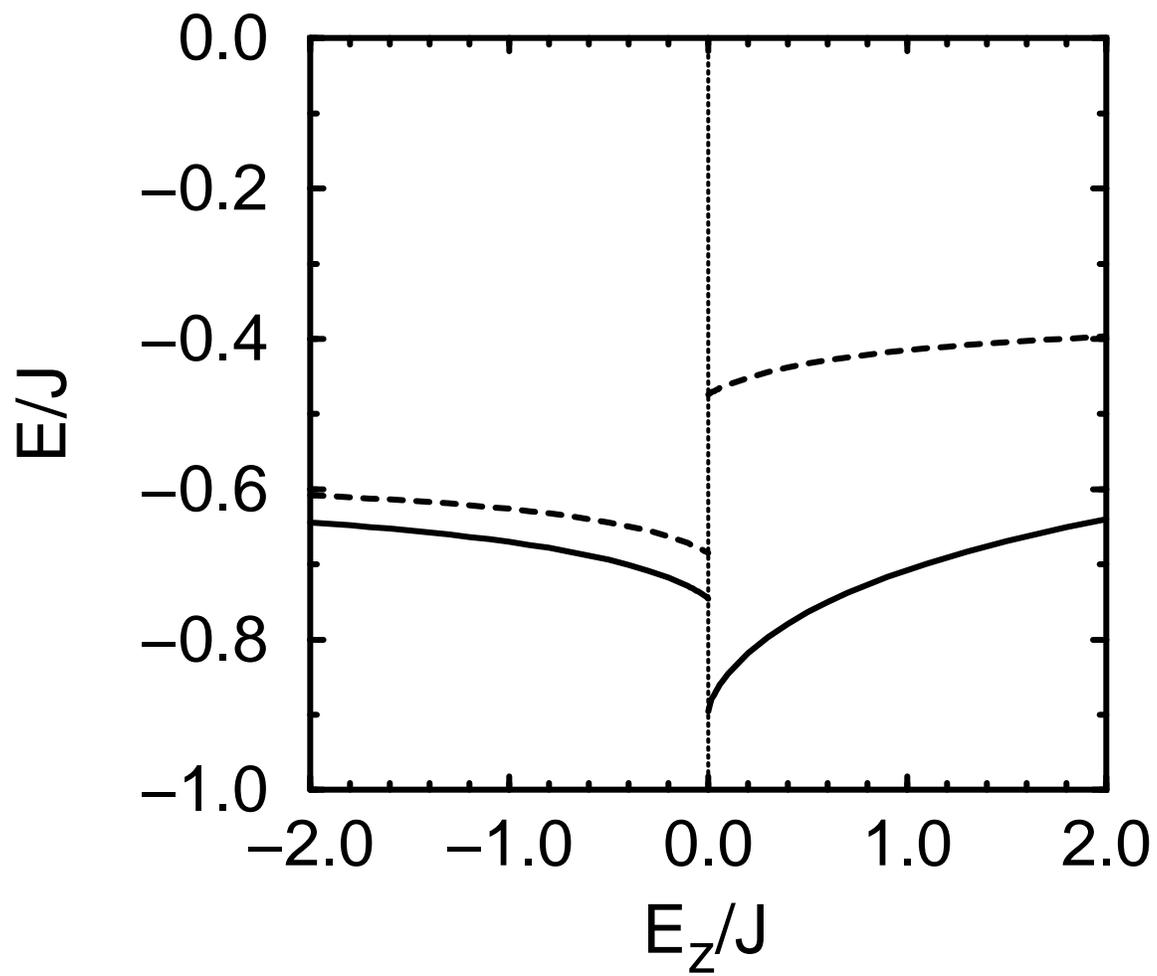
\end{document}